\documentstyle[aps,prl,multicol,epsf]{revtex}
\begin{document}

\begin{abstract}
The theory of fluctuation conductivity for an arbitrary impurity
concentration including ultra-clean limit ($T\tau \gg \sqrt{\frac{T_c}{T-T_c}%
}$) is developed. It is demonstrated that the formal divergency of the
fluctuation density of states contribution obtained previously for the clean
case is removed by the correct treatment of the non-local ballistic electron
scattering. We show that in the ultra-clean limit the density-of-states
quantum corrections are canceled by the Maki-Thompson term and only
classical paraconductivity remains.
\end{abstract}

\begin{center}
{\bf Strong compensation of the quantum fluctuation corrections in clean
superconductor}

D.V.Livanov, G.Savona

{\it Department of Theoretical Physics, Moscow State Institute of Steel and
Alloys, Leninski pr. 4, Moscow 117936, Russia}

A.A.Varlamov

{\it Istituto Nazionale di Fisica della Materia(INFM)-Unit\`{a} ''Tor
Vergata'',}

{\it Dipartimento di Scienze e Tecnologie Fisiche ed Energetiche,\
Universit\`{a} di Roma ''Tor Vergata'', via di Tor Vergata, 00133 Roma, Italy%
}
\end{center}

\vskip 0.5truecm

\vskip 0.5truecm


{\bf \ }As it is well known, the first order fluctuation corrections to
conductivity in the vicinity of superconducting transition are presented by
the Aslamazov-Larkin (AL), Maki-Thompson (MT) and density of states (DOS)
contributions. First one has the simple physical meaning of the direct
charge transfer by the fluctuation pairs themselves and can be easily
derived from the phenomenological time dependent Ginzburg-Landau equation 
\cite{Dorsey93}. In this sense it is a result characteristic for the
classical electron theory, while Maki-Thompson and DOS contributions have
the purely quantum origin and can be calculated in the frameworks of the
microscopic approach only \cite{VBML99}.

The character of the electron scattering plays a very special role for the
manifestation of fluctuation effects. In the BCS theory of superconducting
alloys the only criterion of the metal purity exists: it is the ratio
between the Cooper pair ''size'' (zero temperature coherence length of pure
metal $\xi _0$) and the electron mean free path $\ell .$ If the alloy is
dilute ($\ell \gg \xi _0$) the Cooper pairs motion is ballistic and
impurities do not manifest themselves in superconductor properties. In the
opposite case $\ell \ll \xi _0$ the Cooper pairs motion has the diffusive
character and the role of the effective Cooper pair size plays the
renormalized coherence length $\xi ^{^{\prime }}=\sqrt{\ell \xi _0}.$ The
relative magnitude of fluctuation effects, which is determined by the
Ginzburg-Levanyuk number, is proportional to $(a/\xi )^n$ ($a$ is an
interatomic distance and $n>0$ depends on the effective dimensionality of
the electron spectrum) and it grows for impure systems.

Dealing with the superconductor electrodynamics in fluctuation regime it is
necessary to remember that in the vicinity of the critical temperature the
role of fluctuation Cooper pair effective size plays the Ginzburg-Landau
coherence length $\xi _{GL}(T)=\xi _0/\sqrt{\varepsilon }$ (where the
reduced temperature $\varepsilon =(T-T_c)/T_c$). So the case of dilute metal
($\ell \gg \xi _0$) in the vicinity of the transition could be formally
subdivided on clean, which is still local ($\xi _0\ll \ell \ll \xi _{GL}(T)$%
) and ultra-clean, non-local ($\xi _{GL}(T)\ll \ell $) limits. In terms of
the used in the theory of disordered alloys parameter $T\tau $ the same
three domains can be written down as $T\tau \ll 1;$ $1\ll T\tau \ll 1/\sqrt{%
\varepsilon }$ and $1/\sqrt{\varepsilon }\ll T\tau .$ (We use units $%
k_B=\hbar =c=1$). The latter case was rarely discussed in literature \cite
{BM90,RV94,AG95} in spite of the fact that it becomes of the first
importance already for metals of very modest purity, let us say, $T\tau
\approx 10$. Really, in this case the condition $T\tau \geq 1/\sqrt{%
\varepsilon }$ read for the reduced temperature as $10^{-2}\leq \varepsilon
\ll 1$ practically covers all experimentally accessible range of
temperatures for the fluctuation conductivity measurements. What concerns
the usually considered local clean case ($1\ll T\tau \ll 1/\sqrt{\varepsilon 
}$) for chosen value $T\tau \approx 10$ it would not have any range of
applicability: indeed, the equivalent condition for the allowed temperature
interval $\varepsilon \ll 1/(T\tau )^2$ almost contradicts to the $2D$
thermodynamical Ginzburg-Levanyuk criterion of the mean field approximation
applicability ($Gi\sim \frac{T_c}{E_F}\ll \varepsilon $). Moreover, as it is
well known, for transport coefficients the high order corrections become to
be comparable with the mean field results much before than for
thermodynamical ones, namely at $\varepsilon \sim \sqrt{Gi}$ \cite{LO73,VD83}%
. So in practice one can speak about the dirty, intermediate or ultra-clean
cases but not about the clean one.

We will restrict our consideration by the most interesting case of 2D
electron spectrum, relevant to the high temperature superconductors. As it
is known, the classic 2D AL contribution turns out to be independent on the
electron mean free path $\ell $ at all \cite{AL68}: 
\begin{eqnarray}
\delta \sigma _{AL}^{(2)}=\frac{e^2}{16}\frac 1\varepsilon  \label{ALloc}
\end{eqnarray}

Anomalous Maki-Thompson contribution, being induced by the pairing on the
Brownian diffusive trajectories \cite{VBML99}, naturally depends on $T\tau ,$
but its form changes only when $\ell \sim \xi _{GL}(T)$ ($T\tau \sim 1/\sqrt{%
\varepsilon }$). Its calculation, even with the non-local Cooperon vertices
but the standard propagator, in the ultra-clean limit \cite{RV94} leads to
the expression less divergent in its temperature dependence but growing as $%
T\tau \ln (T\tau )$ with the increase of $\ell $ \cite{BM90,RV94}: 
\begin{equation}  \label{MTloc}
{\sigma }_{MT(an)}^{(2)}=\frac{e^2}8\left\{ 
\begin{tabular}{c}
${\frac 1{\varepsilon -\gamma _\varphi }}\ln (\varepsilon /\gamma _\varphi
),T_c\tau \ll 1/\sqrt{\varepsilon }$ \\ 
${\frac{8\pi ^2T\tau }{\sqrt{14\zeta (3)}}\frac 1{\sqrt{\varepsilon }}}\ln
(T_c\tau \sqrt{\varepsilon }),1/\sqrt{\varepsilon }\ll T_c\tau $%
\end{tabular}
\right. ,  \label{mt}
\end{equation}
where $\gamma _\phi =\pi /8T\tau _\phi $ is the inelastic scattering rate.

The analogous problem takes place in the DOS and regular part of MT
contributions: their standard diagrammatic technique calculations lead to
the negative correction \cite{ILVY93} 
\begin{eqnarray}
\sigma _{DOS+MT(reg)}^{(2)}=-\frac{e^2}{4s}\kappa (T\tau )\ln \left( \frac 1%
\varepsilon \right) ,  \label{dos}
\end{eqnarray}
\begin{equation}
\kappa (T\tau )={\frac{{-\psi ^{^{\prime }}({\frac 12}+{\ \frac 1{{4\pi \tau
T}}})+{\frac 1{{2\pi \tau T}}}\psi ^{^{\prime \prime }}({\frac 12})}}{{\pi
^2[\psi ({\frac 12}+{\frac 1{{\ 4\pi \tau T}}})-\psi ({\frac 12})-{\frac 1{{%
4\pi \tau T}}}\psi ^{^{\prime }}({\frac 12})]}}}=\left\{ 
\begin{tabular}{c}
$\frac{56\zeta (3)}{\pi ^4}\approx 0.691,\;T\tau \ll 1$ \\ 
$\frac{8\pi ^2}{7\zeta (3)}(T\tau )^2,\;T\tau \gg 1$%
\end{tabular}
\right.  \nonumber
\end{equation}
evidently divergent when $T\tau \rightarrow \infty .$

In the derivation of these results the local form of the fluctuation
propagator and Cooperons (besides (2)) were used. It is why in view of the
mentioned above peculiarity of ultra-clean limit, the extension of their
validity for $T\tau \gg 1/\sqrt{\varepsilon }\rightarrow \infty $ seems to
be doubtful.

One can notice that at the upper limit of the clean case, when $T\tau \sim 1/%
\sqrt{\varepsilon }$ both DOS and anomalous MT contributions turn out to be
of the same order of value but of the opposite signs. So one can suspect
that in the case of correct procedure of the impurities averaging in the
ultra-clean case the large negative DOS contribution can be cancelled with
the positive anomalous MT one.

The reexamination of all fluctuation corrections of the first order in the
case of the arbitrary impurity concentration including non-local electron
scattering in the ultra-clean superconductor will be the aim of this
communication. The nontrivial cancellation of the contributions, previously
divergent in $T\tau ,$ will be shown. It results in the reduction of the
total fluctuation correction in ultra-clean case to the AL term only.

In purpose to calculate the Cooperon (impurity vertex) $C({\bf q},\epsilon
_1,\epsilon _2)$ and fluctuation propagator $L({\bf q},\omega _\mu )$ (the
two-particle Green function) in the general case of an arbitrary electron
mean free path case one needs the explicit expression for the polarization
operator ${\cal P}({\bf q},\epsilon _1,\epsilon _2),$ which due to the
elasticity of scattering does not contain the frequency summation and for $%
2D $ spectrum has a form: 
\begin{eqnarray}
{\cal P}({\bf q},\epsilon _1,\epsilon _2)=\int \frac{d^2{p}}{(2\pi )^2}G(%
{\bf p+q},\epsilon _1)G({-}{\bf p},\epsilon _2)=\frac{2\pi N(0)\Theta
(-\epsilon _1\epsilon _2)}{\sqrt{v_F^2q^2+(\widetilde{\epsilon }_1-%
\widetilde{\epsilon }_2)^2}},  \label{a2}
\end{eqnarray}
where $\Theta (-x)$ is the Heaviside theta-function, $\widetilde{\epsilon }%
_n=\epsilon _n+\frac 1{2\tau }sgn\epsilon _n,$ $N(0)$ and $v_F$ are the
density of states and the velocity at the Fermi level. Let us stress that
this result was carried out without any expansion over the Cooper pair
center of mass momentum ${\bf q}$ and is valid for an arbitrary $\ell q.$

The standard ladder consideration results in the following expressions for
the Cooperon and fluctuation propagator: 
\begin{equation}
C^{-1}({\bf q},\epsilon _1,\epsilon _2)=1-{\frac{\Theta (-\epsilon
_1\epsilon _2)}{{\tau }\sqrt{v_F^2q^2+(\widetilde{\epsilon }_1-\widetilde{%
\epsilon }_2)^2}}}.  \label{a14}
\end{equation}
and 
\begin{eqnarray}
-[N(0)L({\bf q},\Omega _\mu )]^{-1} &=&\ln \frac T{T_c}+\sum_{n=0}^\infty
\left\{ \frac 1{n+1/2}\right.  \nonumber \\
&&\ \left. -\frac 1{\sqrt{\left( n+\frac 12+\frac{\Omega _\mu }{4\pi T}+%
\frac 1{4\pi T\tau }\right) ^2+\frac{v_F^2{\bf q}^2}{16\pi ^2T^2}}-\frac 1{%
4\pi T\tau }}\right\}  \label{genpro}
\end{eqnarray}
Near $T_c$ $\ln \frac T{T_c}\approx \varepsilon $ and for the local limit,
when just small momenta $\ell q\ll 1$ are involved in the final
integrations, the Eqs. (\ref{a14}) and (\ref{genpro}) can be expanded over $%
v_Fq/\max \{T,\tau ^{-1}\}$ and they are reduced to the well known local
expressions.

The Feynman diagrams which contribute to conductivity in the first order of
perturbation theory on electron-electron interaction in Cooper channel are
presented in Fig.1. Let us start from the discussion of the Maki-Thompson
contribution (diagram 6). We restrict our consideration by the vicinity of
the critical temperature, where, for the most singular in reduced
temperature contribution, static approximation is valid. It means that
Cooper pair bosonic frequency can be assumed $\Omega _\mu =0$.

Using the general expressions for Cooperons and propagator (\ref{a14}), (\ref
{genpro}) after the integration over electronic momentum, one can find: 
\begin{eqnarray}
Q^{(6)}\left( \omega _\upsilon \right) &=&-4\pi
N(0)v_F^2e^2T^2\sum_{\varepsilon _n}\int \frac{d^2{\bf q}}{(2\pi )^2}L({\bf q%
},0)\times  \label{MT1} \\
&&\left[ {\cal M}\left( \stackrel{\sim }{\epsilon }_n,\stackrel{\sim }{%
\epsilon }_{n+\nu },{\bf q}\right) +{\cal M}\left( \stackrel{\sim }{\epsilon 
}_{n+\nu },\stackrel{\sim }{\epsilon }_n,{\bf q}\right) \right] ,  \nonumber
\end{eqnarray}
where 
\begin{equation}
{\cal M}\left( \alpha ,\beta ,{\bf q}\right) =\frac{\ R_q(2\alpha {\bf )}%
R_q(\alpha +\beta {\bf )}-\Theta (\alpha \beta )R_q(2\alpha {\bf )}%
R_q(2\beta {\bf )}}{(\beta -\alpha )^2\left( R_q(2\alpha {\bf )}-\frac 1\tau
\right) \left( R_q(2\beta {\bf )}-\frac 1\tau \right) R_q(\alpha +\beta {\bf %
)}},  \label{MT2}
\end{equation}
$R_q{\bf (}x{\bf )}=\sqrt{x^2+v_F^2{\bf q}^2}$.

The analogous consideration of the main in the clean case DOS diagrams 2 and
4 leads to: 
\begin{eqnarray}
\ Q^{(2+4)}\left( \omega _\upsilon \right) &=&4\pi
N(0)v_F^2e^2T^2\sum_{\varepsilon _n}\int \frac{d^2{\bf q}}{(2\pi )^2}L({\bf q%
},0)\times  \label{DOS1} \\
&&\ \left[ {\cal D}(\stackrel{\sim }{\epsilon }_n,\stackrel{\sim }{\epsilon }%
_{n+\nu },{\bf q})+{\cal D}(\stackrel{\sim }{\epsilon }_{n+\nu },\stackrel{%
\sim }{\epsilon }_n,{\bf q})\right] ,  \nonumber
\end{eqnarray}
with 
\begin{eqnarray*}
{\cal D}(\alpha ,\beta ,{\bf q}) &=&(\beta -\alpha )^2\left( R_q{\bf (}%
2\alpha {\bf )-}\frac 1\tau \right) ^2\times \\
&&\left[ \frac{R_q^2{\bf (}2\alpha {\bf )+}2\alpha (\alpha -\beta )}{R_q{\bf %
(}2\alpha {\bf )}}-\frac{\Theta \left( \alpha \beta \right) R_q^2{\bf (}%
2\alpha {\bf )}}{\left( R_q{\bf (}\alpha +\beta {\bf )-}\frac 1\tau \right) }%
\right]
\end{eqnarray*}

One can see that each of expressions for $Q^{(6)}\left( \omega _\upsilon
\right) $ and $Q^{(2+4)}\left( \omega _\upsilon \right) ,$ in accordance
with \cite{ILVY93,DKVBL93}, in the limit $T\tau \rightarrow \infty $
presents itself the Loran series of the type $C_{-2}(T\tau )^2+C_{-1}(T\tau
)+C_0+C_1(T\tau )^{-1}+...$ . The careful expansion of the sum of
expressions (\ref{MT1}) and (\ref{DOS1}) in the series of such type leads to
the exact cancellation of all divergent contributions and even to the
coefficient $C_0=0$. In result, the leading order of the sum of MT and DOS
contributions in the limit of $T\tau \gg 1$ turns out to be $C_1(T\tau
)^{-1} $ and it disappears in the non-local limit. The results of numerical
calculation of $Q^{(6)}\left( \omega _\upsilon \right) +$ $Q^{(2+4)}\left(
\omega _\upsilon \right) $ as the function of $T\tau $ according to Eqs. (%
\ref{MT1}) and (\ref{DOS1}) are presented in Fig. 2 for different
temperatures. One can convince himself in the rapid decrease of this sum
with the increase of $T\tau $.

The remained four diagrams among the first order fluctuation corrections to
conductivity (see, for example, the Fig.9 in the review article \cite{VBML99}%
) are negligible in the vicinity of $T_c$. The similar consideration of the
remaining two DOS-like diagrams (3) and (5) gives 
\begin{eqnarray}
Q^{(3+5)}(\omega _\upsilon ) &=&4\pi N(0)v_F^2e^2T^2\sum_{\varepsilon
_n}\int d^2{\bf q}L({\bf q},0)\times  \nonumber \\
&&\left[ {\cal K}(\stackrel{\sim }{\epsilon }_n,\stackrel{\sim }{\epsilon }%
_{n+\nu },{\bf q})+{\cal K}(\stackrel{\sim }{\epsilon }_{n+\nu },\stackrel{%
\sim }{\epsilon }_n,{\bf q})\right] ,  \label{DOSi1}
\end{eqnarray}
where 
\begin{equation}
{\cal K}(\alpha ,\beta ,{\bf q})=\frac{2\beta \Theta (-\alpha \beta )}{%
(\beta -\alpha )\left( R_q(2\alpha {\bf )}-\frac 1\tau \right) ^2R_q(2\alpha 
{\bf )}}  \nonumber  \label{DOSi2}
\end{equation}
Evaluation of Eq.(\ref{DOSi1}) demonstrates that for $T\tau \gg 1$ the final
contribution of diagrams 3 and 5 does not contain $\tau -$dependence, and is
less ($\propto \ln 1/\varepsilon $) singular if compared with
paraconductivity, in spite of the fact that each of diagrams 3 and 5
contains the divergent Loran term $\propto T\tau $ which cancel each other.

So one can see that the DOS term divergence $\propto (T\tau )^2,$ found
before for clean case \cite{ILVY93,DKVBL93} , has a restricted validity and
can not be extended to $T\tau \rightarrow \infty $. In the limit of
defectless superconductor the total DOS+MT contribution is proportional to $%
\ln 1/\varepsilon $ and is independent on $T\tau .$

Finally let us turn to the discussion of the AL contribution. In this case,
as it is well known, even in the vicinity of $T_c$ we cannot restrict
ourselves by the static approximation and analytical continuation over the
external frequency has to be accomplished. The diagram 1 at Fig.1 represents
the Aslamazov-Larkin contribution: 
\begin{equation}
Q^{AL}(\omega _\upsilon )=\frac{e^2}{2\pi i}\int \frac{d^2{\bf q}}{(2\pi )^2}%
\oint dz\coth \frac z{2T}B^2({\bf q,}z,\omega _\upsilon )L({\bf q},-iz)L(%
{\bf q},-iz+\omega _\upsilon ),  \label{AL1}
\end{equation}
where the three Green function blocks have to be calculated with the
non-local Cooperons and the expression for the non-local propagators have to
be used. This program is hard to be realized in the general form and at
present has been tried to be solved with different approximations in the set
of papers \cite{RVV90,AHL95,ERS99,Ax99}. We are interested here to study
the effect of non-locality on the AL contribution in the first hand
sacrificing the $ac$ effects ($\omega \neq 0$) and being in the vicinity of
the transition $\varepsilon \ll 1.$. So we omit the $z,\omega _\nu $
dependence of the Green functions block

\begin{equation}
B({\bf q,}z=0,\omega _\upsilon =0)=-4\pi N(0)T\upsilon
_F^2q\sum\limits_{n=0}^\infty \frac 1{\left( \sqrt{4\widetilde{\epsilon _n}%
^2+\upsilon _F^2{\bf q}^2}-\frac 1\tau \right) ^2\left( \sqrt{4\widetilde{%
\epsilon _n}^2+\upsilon _F^2{\bf q}^2}\right) }  \label{AL2}
\end{equation}
and evaluate the AL contribution in this approximation numerically. These
calculations show that the temperature dependence of paraconductivity turns
out to be close to the classical 2D $\varepsilon ^{-1}$regime. It is
necessary to mention that relatively far from the transition, where our
approximation strictly speaking is already not applicable, the calculated
curve lies somewhat lower of the AL theory prediction in accordance with the
short wave-length fluctuation calculations and experimental findings \cite
{VBML99}. We note, that although Eqs. (\ref{AL1}), (\ref{AL2}) contain
dependence on $\tau $, the paraconductivity is turned out to be $\tau $%
-independent in the entire range of parameter $T\tau $, analogous to the
local 2D result (\ref{ALloc}).

Let us discuss the results obtained. First of all it is necessary to stress
the observed strong cancellation of the DOS and MT contributions, which were
found previously, within the local fluctuation theory, to be divergent in
the limit $\tau \rightarrow \infty $ \cite{ILVY93,DKVBL93}. As we have
demonstrated, the correct account of non-local scattering processes in the
ultra-clean limit results in the impurity independent, logarithmic in
reduced temperature, contribution negligible in comparison with the more
singular AL one.

Let us remind that the fluctuation conductivity in the limiting case $\tau
=\infty $ and for the non-zero frequency of the external electromagnetic
field was studied in Ref. \cite{RVV90}, where the similar problem of the $%
\omega ^{-2}$-divergence( instead of $\tau ^2$ in our case) of the
contributions from different DOS and MT-like diagrams aroused. The sum of
all relevant diagrams nevertheless was found to be regular and proportional
to $\omega .$ Moreover, the sum of all DOS and MT diagrams (which correspond
to diagrams 6, 2 and 4 in Fig. 1) was shown to be zero in the case $\tau
=\infty .$ In the current publication we have confirmed this statement
studying the more general case $\omega \rightarrow 0,\tau \rightarrow \infty 
$ with $\omega \tau \rightarrow 0$ and convincing ourselves that for the
one-electron type DOS and MT fluctuation processes the final result does not
depend on the order of the $\lim_{\omega \rightarrow 0,\tau \rightarrow
\infty }(\sigma ^{DOS}+\sigma ^{MT})$ calculation. We have evaluated the
explicit dependence of the overall fluctuation conductivity on the parameter 
$T\tau $ and have demonstrated its regular character.

What concerns the AL contribution, the careful investigation of its clean
limit was done in \cite{AHL95} by means of the analysis of the
paraconductivity diagram structure in the coordinate representation. It was
shown that the electric field does not interact directly with the
fluctuation Cooper pairs, but it produces the effect on the quasiparticles
forming these pairs only. The characteristic time of the change of the
quasiparticle state is of the order of $\tau $ . Consequently the
one-particle Drude type conductivity in $ac$ field, as it is well known, has
the first order pole. What concerns the AL paraconductivity, due to the
above mentioned reasons, the pole in it was found to be of the second order 
\cite{AHL95}:

\begin{equation}
\sigma _{AL}(\omega )=\frac{\sigma _{AL}^{dirty}}{(1-i\omega \tau )^2}
\label{Dr}
\end{equation}
In spite of this difference one can see that the AL conductivity, like, the
Drude one, vanishes at $\omega \neq 0,\tau \rightarrow \infty $ because in
the absence of impurities the interaction of electrons does not produce any
effective force acting on the superconducting fluctuations, while the $dc$
paraconductivity conserves its usual $\tau -$independent form.

In the present paper we have approached to the same problem of the
investigation of the AL contribution in clean metal studying the general
non-local case in q-space and have shown the independence of the $dc$
paraconductivity on the material purity.

It is necessary to stress that the non-local forms of the Cooperon and
fluctuation propagator have to be accounted not only for the ultra-clean
case but in every problem where the relatively large bosonic momenta are
involved: account for the dynamical and short wavelength fluctuations beyond
the vicinity of critical temperature, the effect of relatively strong
magnetic fields on fluctuations and weak localization corrections etc.
Recently such approach was developed in the set of studies of the DOS
fluctuation effects \cite{BM90,RV94,ERS99} and the efforts to apply it to
the complete microscopic calculation of the magnetoconductivity for
arbitrary temperatures and fields is undertaken in \cite{Ax99}.

Authors are grateful to J.Axnas and C.Castellani for valuable discussions.
This work was accomplished in the frameworks of the INTAS Grant \# 96-0452.

\newpage\ 

{\bf Figure Captions}

Fig.1. Feynman diagrams for the leading-order contributions to fluctuation
conductivity. Wavy lines are fluctuation propagators, thin solid lines with
arrows are impurity-averaged normal-state Green's functions, shaded
semicircles are vertex corrections arising from impurities, dashed lines
with central crosses are additional impurity renormalizations and shaded
rectangles are impurity ladders. Diagram 1 is the Aslamazov-Larkin term;
diagrams 2-5 arise from the corrections to the normal state density of
states in the presence of impurity scattering; diagram 6 is the
Maki-Thompson term.

Fig. 2. The illustration of the decrease of the sum of DOS and MT
contributions with the increase of the mean free path for the different
values of reduced temperature: $\varepsilon =0.001;0.01;0.1;0.2.$

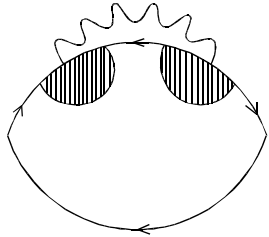
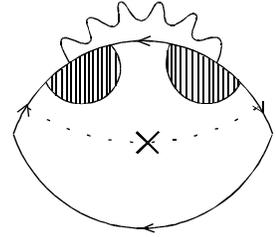
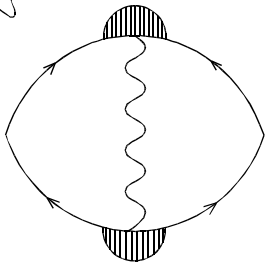
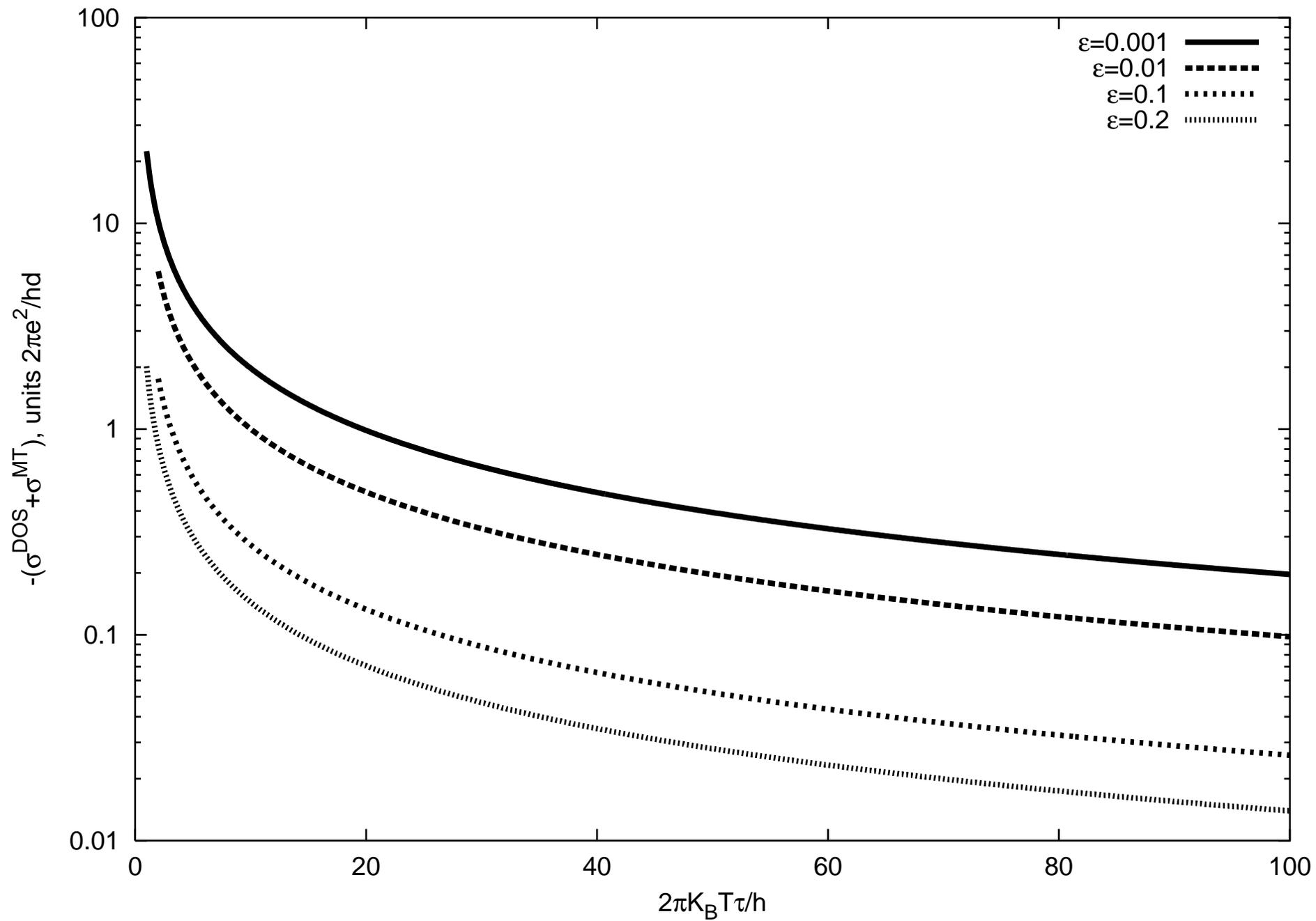
\end{document}